\documentclass[num-refs, twocolumn]{wiley-article}

\usepackage{siunitx}
\usepackage{dirtytalk}
\usepackage{booktabs}
\usepackage{tabularx}

\papertype{Research Article}

\title{Do Students Learn Better Together? Teaching Design Patterns and the OSI Model with the Aronson Method}

\author[1]{Daniel San Martín}
\author[1]{Carlos Manzano}
\author[2]{Valter Vieira de Camargo}

\affil[1]{Escuela de Ingeniería, Universidad Católica del Norte, Coquimbo, Chile}
\affil[2]{Departamento de Computação, Universidade Federal de São Carlos, São Carlos, SP, Brazil}

\corraddress{Daniel San Martín, Escuela de Ingeniería, Universidad Católica del Norta, Coquimbo, Chile}
\corremail{daniel.sanmartin@ucn.cl}

\fundinginfo{Fondo de Desarrollo de Proyectos Docentes de Pregrado 2024, DGPRE, UCN, N°181/2024}

\runningauthor{San Martin et al.}

\begin{document}

\begin{frontmatter}
	\maketitle
	
	\begin{abstract}
	Abstract concepts like software design patterns and the OSI model often pose challenges for engineering students, and traditional methods may fall short in promoting deep understanding and individual accountability. This study explores the use of the Aronson Jigsaw method to enhance learning and engagement in two foundational computing topics. The intervention was applied to two 2025 cohorts, with student progress measured using a Collaborative Learning Index derived from formative assessments. Final exam results were statistically compared to previous cohorts. While no significant correlation was found between the index and final grades, students in the design patterns course significantly outperformed earlier groups. Networks students showed more varied outcomes. Qualitative trends point to cognitive and metacognitive gains supported by peer teaching. The Jigsaw method encourages collaborative engagement and may support deeper learning. Future work will explore the integration of AI-based feedback systems to personalize instruction and further improve learning outcomes.
	\keywords{collaborative learning, jigsaw method, software design patterns, OSI model, engineering education, student engagement}
	\end{abstract}
\end{frontmatter}

\section{Introduction} \label{intro}
The \textit{2030 Agenda for Sustainable Development}, adopted by the United Nations General Assembly, outlines a global plan centered on human well-being, environmental protection, and the promotion of inclusive societies \citep{UN2015}. Within this framework, Sustainable Development Goals 4 and 5 emphasize inclusive and equitable education, lifelong learning, and the active participation of women in decision-making. In particular, Target 4.7 promotes the acquisition of skills to advance sustainable development from a gender equity perspective \citep{UN2015,UNESCO2017}.

Current pedagogical trends increasingly advocate for strategies that connect academic content to real-world contexts through participatory and collaborative methodologies \citep{LeyvaMoral2016, Morales2020, Seymour2025}. In this context, \textit{cooperative learning (CL)} has emerged as a strategy supported by both theoretical and empirical research \citep{Johnson2002, Johnson2017, Pujolas2009, Canabate2020}, promoting the development of academic and social skills simultaneously \citep{Slavin1995, Johnson2009, Johnson1993, Velazquez2015, Haerens2011}.

Despite its potential, group work practices in higher education often rely on unstructured collaboration, which may fail to ensure equitable participation or improved learning outcomes \citep{Kagan1992, AndreuAndres2015}. This issue is especially relevant in fields such as computer science and engineering, where academic performance is frequently low \citep{Lazaro2017, JONES2022}, and courses such as design patterns \citep{Abdul2007, Holubek2022} and computer networks \citep{Abdrabou2021, Rahman2023} often exhibit high failure rates. These challenges are intensified by traditional instructional models that emphasize individual competition, contributing to student demotivation and hindering meaningful learning.

Moreover, some studies have shown that in more diverse educational contexts, gender differences can emerge in participation and perceptions of cooperative learning. Although this was not a central focus of the present study due to the low number of female students in the analyzed courses, we acknowledge that such dynamics may influence outcomes in more heterogeneous environments \citep{Jordan2001, Gupta2014, Cigarini2020}. These inequalities can persist even when using cooperative strategies if the methods are not implemented effectively, potentially undermining their intended benefits \citep{Yang2019, Berenbaum2019}.

These challenges underscore the need to reflect on how cooperative strategies are applied in technical and heterogeneous learning environments. This study analyzes the implementation of the \textit{Aronson Jigsaw technique} as a cooperative learning strategy to improve academic performance and promote equitable participation in the teaching of design patterns and OSI model layers in computer science courses. In this methodology, each topic was divided into sections assigned to individual students, who became experts on their portion and then taught it to their peers \citep{Aronson1997}. This structure fosters positive interdependence and individual accountability \citep{Dyson2002, SuarezCunqueiro2017}. Heterogeneous groups were formed based on ability, gender, and academic background, following best practices for equitable group design \citep{Putnam1999, Kagan1992}.

In addition to assessing academic outcomes, this study also examined students’ perceptions of the technique, particularly regarding coordination, fairness, and their sense of progress. This approach evaluates both the pedagogical effectiveness of the Jigsaw method and its potential to enhance learning in technical university settings. Although gender differences are not analyzed in depth, the study recognizes that future research could address this dimension in more representative settings.

The article is structured as follows. Section~\ref{relwork} reviews background and related work. Section~\ref{context} presents the empirical setting. Sections~\ref{caseA} and~\ref{caseB} describe two teaching cases using the Aronson method. Section~\ref{analysis} analyzes student outcomes. Section~\ref{discussion} provides a broader interpretation. Section~\ref{limitations} presents the threats of validity and Section~\ref{conclusion} concludes with final remarks and future directions.

\section{Background \& Related Work} \label{relwork}
Within the framework of \textit{Cooperative Learning} (CL), various methods and techniques are employed. One of the ten most widely used techniques today, according to Johnson and Johnson \citep{Johnson2002}, is the \textit{Aronson Puzzle} (AP) technique \citep{Aronson1978}. This statement is reinforced by \citet{Velazquez2015}, who argues that AP is one of the preferred methodologies among instructors, while \citet{FernandezRio2017} considers it the best-known and most widely implemented CL technique, despite comparisons made by some authors with similar methodologies, such as Problem-Based Learning. Nonetheless, \citet{LeyvaMoral2016} emphasize that AP possesses its own identity as a didactic technique.

The AP method consists of dividing the learning content into multiple sections. Each member of a group specializes in one of these parts and later teaches it to their peers, who in turn have become experts in other \textit{pieces of the puzzle} \citep{Aronson1997}. In this way, each student is responsible for mastering a specific subtopic and becomes an expert on that subject. This information-sharing process gives each student an indispensable role within the group, making their contribution essential for achieving the common goal. Consequently, AP fosters positive interdependence and individual accountability among group members \citep{Dyson2002, SuarezCunqueiro2017}.

Widely recognized among active learning methodologies \citep{Johnson2002, Velazquez2015, Aronson1978, FernandezRio2017}, the AP technique proposes a structured division of content, whereby each student specializes in a \textit{piece of the puzzle} and then teaches it to their peers \citep{Aronson1997}. This technique has been successfully applied in disciplines such as computer science \citep{Anguas2007}, medi\-cine \citep{Buhr2014, Renganathan2020}, dentistry \citep{SuarezCunqueiro2017}, and sport sciences \citep{Acosta2019}, fostering both positive interdependence and shared responsibility for learning \citep{Dyson2002, SuarezCunqueiro2017}. However, research has also highlighted certain challenges: limited student perception of learning progress \citep{Bratt2008}, difficulties in integrating content taught by peers \citep{FernandezRio2017}, lack of faculty preparation, and resistance to the shift away from traditional teacher-centered roles \citep{Schoenecker1997, Ellison1994}. Nevertheless, AP is widely recognized for promoting active participation, interpersonal relationships, and a sense of group belonging \citep{LeyvaMoral2016, FernandezRio2017}.

\citet{Bratt2008} also acknowledges valuable features of cooperative learning, such as the strengthening of group identity and positive attitudes toward learning. However, he notes that the technique does not always generate a clear perception of progress in knowledge acquisition. Along the same lines, \citet{Schoenecker1997} stresses the importance of understanding AP’s formative benefits and avoiding its interpretation as a mere sum of parts. Previous studies have identified difficulties in implementation, including lack of coordination in assigned tasks and individual irresponsibility within groups \citep{Ellison1994}.

\citet{FernandezRio2017} argues that both faculty and students are often unprepared to implement AP correctly. Among the factors that hinder its effectiveness, the author cites inefficiencies in integrating content taught by group \textit{experts}, as well as resistance to changes in the teacher’s role and perceived loss of control. In AP, content mastery depends directly on peer interaction, making each member’s performance crucial to reaching shared learning objectives. This process fosters interpersonal relationships and requires students to adapt their personalities to the dynamics of cooperative learning environments. Nevertheless, research exploring gender-based differences in students’ perceptions within cooperative learning environments generated by the AP technique remains limited.

These implementation challenges are particularly relevant when addressing complex and abstract topics such as design patterns and the OSI model—core components of computer engineering education. Design patterns are reusable solutions to common problems in object-oriented programming. They facilitate communication among developers and promote software maintainability, but require a deep understanding of the context in which they should be applied. According to \citet{Ferrandis2021}, many students struggle to identify when and how to apply a specific pattern, resulting in a gap between theoretical knowledge and practical implementation. To address this issue, the author developed an interactive web tool that provides adaptive, visual examples, thereby supporting students’ understanding of these abstract concepts.

The OSI (Open Systems Interconnection) model, in turn, is a reference architecture that divides the network communication process into seven layers, from the physical to the application level. Although essential for understanding how computer networks function, its hierarchical structure and the lack of applied examples make it difficult for students to grasp. A study by \citet{UNICEN2019} underscores the importance of breaking down the OSI model into more comprehensible processes that connect theory with practice. Similarly, \citet{Sarmiento2019} highlights the use of real-world problems based on TCP/IP protocols to help students contextualize the behavior of each layer.

In both cases, design patterns and the OSI model, the literature agrees on the need to incorporate active methodologies, gamified strategies, and interactive digital resources that encourage student engagement and understanding, while reducing the high failure rates in these essential core courses in the computing curriculum.

\section{Context \& Empirical Basis} \label{context}
This section outlines the pedagogical setting and implementation details for two undergraduate courses in which the Aronson Puzzle technique was applied: \textit{Software Engineering}, focusing on design patterns, and \textit{Computer Networks}, focusing on the OSI model. The empirical basis includes the structure of the sessions, student grouping, instructional materials, and the integration of cooperative learning tools such as Socrative, mental maps, and peer evaluation instruments.

\subsection{Implementation in Software Engineering: Design Patterns}

This teaching case was implemented during a regular semester in a third-year undergraduate Pattern Software and Programming course at a Chilean university. The intervention addressed the difficulties students face when learning the abstract and technical nature of the GoF design patterns. To tackle this issue, the Aronson Puzzle (Jigsaw) cooperative learning technique was applied across three thematic sessions: creational, structural, and behavioral patterns.

The course had a total of 20 students, all of whom participated in the activity as part of the mandatory curriculum. Instructional materials included teacher-designed slides and study guides. Students were divided into 5 base groups of 4 members each.

Before the intervention, an introductory session was held to present the methodology, explain the expectations of the Aronson Puzzle technique, and conduct a basic characterization of the student cohort. The purpose was to familiarize students with the dynamics of cooperative learning and assess baseline conditions for participation.

In the first session focused on creational patterns, students were randomly assigned to base groups and numbered from 1 to 4. Each number corresponded to a specific creational pattern (e.g., Singleton, Factory Method, or Builder). Students first read the assigned material individually and then reorganized into expert groups (all students with the same number). In expert groups, they discussed their pattern collaboratively. After a designated time, they returned to their original base groups and each student taught their assigned pattern to their peers.

In the following class, students completed an individual formative quiz through Socrative to assess comprehension. Afterwards, they presented group-created mental maps of the assigned pattern category. The instructor facilitated these presentations with probing questions. The session concluded with students completing a self- and peer-evaluation form focused on participation, clarity of explanation, and group coordination.

This cycle was repeated for structural patterns in the subsequent two sessions, and again for behavioral patterns across two additional sessions. The structural patterns included: Proxy, Adapter, Facade, and Decorator. The behavioral patterns included: Strategy, Observer, Template Method, and Visitor. Each session followed the same structure, ensuring consistency across topics and reinforcing the cooperative learning routine.

At the end of the instructional unit, a satisfaction survey was administered to gather student perceptions of the technique, including dimensions such as clarity, engagement, collaborative atmosphere, and perceived learning gains.

The 2025 cohort that experienced this methodology was compared to a 2023-2024 control group that followed traditional instruction. Students in the 2025 group also completed peer and self-evaluation rubrics and constructed conceptual maps. A composite assessment indicator, the Collaborative Learning Index (CLI), was introduced. This metric aggregated individual quiz scores, group performance, and self-assessed participation.

\subsection{Implementation in Computer Networks: OSI Model}

The Computer Networks course~(\cite{kurose2020}) was implemented during a regular semester in an eighth-year undergraduate course. The intervention was designed to address students' difficulties in understanding the abstract and technical nature of the layers of the OSI (Open Systems Interconnection) model.

The course had a total of 15 students, who participated in the activity as part of the required curriculum. The teaching materials included slides and study guides designed by the instructor. Students were divided into three main groups of 5 members each.

Prior to the intervention, an introductory session was held. Its purpose was to introduce Aronson's Jigsaw technique, define expectations for cooperative learning, and obtain a basic characterization of the student group to assess the initial conditions for participation.

The Aronson Jigsaw cooperative learning technique was applied, structured in three thematic sessions, each focusing on a group of OSI layers, which facilitated the understanding of their interaction and function:

\begin{itemize}

    \item \textbf{Session 2:} The transport and network layers (intermediate layers) were addressed, focusing on five key data transmission protocols: TCP (Transmission Control Protocol), UDP (User Datagram Protocol), IP/ICMP (Internet Control Message Protocol), EIGRP (Enhanced Interior Gateway Routing Protocol), and OSPF (Open Shortest Path First). Each student became the expert for one protocol.

    \item \textbf{Session 3:} The data link and physical layers (lower layers) were examined, including the operation of five core technologies and protocols: Ethernet, CSMA/CD (Carrier Sense Multiple Access with Collision Detection), PPP (Point-to-Point Protocol), Token Ring, and physical layer elements (such as transmission media, hardware components, and standards that define electrical or optical signals).
\end{itemize}

This adaptation allows students to break down the complexity of the OSI model, focusing on the inherent abstraction of each layer and how they contribute to the overall functioning of the network.

In the first session, students were randomly assigned to three main groups, each representing a set of upper-layer protocols of the OSI model selected by the instructor (e.g., Group 1 was assigned HTTPS, SMTP, SSH, RPC, SSL, and FTP). At this point, students read about each protocol individually. Then, following Aronson's guidelines, students broke up and reconvened in new groups called experts to collaboratively discuss their knowledge of each of the studied protocols. Finally, they returned to their original (main) groups to share what they had learned with their peers. Afterward, students took an individual formative test using Socrative software\footnote{For more information, visit: \url{https://www.socrative.com/}} to assess their understanding. They also presented mind maps created by the group for their assigned OSI protocol.

This procedure was repeated for subsequent sessions related to the transport/network layers and for the link and physical layers. Each session followed the same structure, ensuring consistency between topics and reinforcing the cooperative learning routine.

At the end of the teaching unit, a satisfaction survey was administered to gather student perceptions of the technique, including dimensions such as clarity, engagement, collaborative environment, and perceived learning achievements.

The 2025 cohort of the computer networking course that underwent this methodology was compared to a 2023-2024 control group that followed traditional instruction. Students in the 2025 cohort also completed self- and peer-assessment rubrics and created concept maps. A composite assessment indicator, the Collaborative Learning Index (CLI), was introduced. This metric aggregated individual quiz scores, group performance, and self-assessed engagement.

\section{Teaching Case A: Software Design Patterns with Aronson} \label{caseA}
This section presents the implementation of the Aronson cooperative learning technique in the course \textit{Patrones de Software y Programación}, emphasizing its pedagogical rationale, instructional design, and preliminary impact. The context, course structure, and characteristics of the student group are detailed to provide a comprehensive foundation for evaluating the effectiveness of this teaching strategy.

\subsection{Course Description and Context}

The course Patrones de Software y Programación (Software Patterns and Programming) is a core component of the fifth-semester curriculum for students enrolled in the Ingeniería Civil en Computación e Informática and Ingeniería en Tecnologías de Información programs. It carries 5 SCT credits (Sistema de Créditos Transferibles), a Chilean standard for quantifying student workload based on the time required to achieve specific learning outcomes in higher education. The course is designed to develop students’ abilities to design, implement, and evaluate robust software systems, with a strong emphasis on the use of established design principles and practices rooted in Software Engineering.

A central focus of the course is the study of GoF (Gang of Four) design patterns, which are grouped into three categories: creational, structural, and behavioral. Students learn to apply patterns such as Singleton, Factory Method, Adapter, Decorator, Strategy, and Observer, among others, in order to create maintainable and scalable software. These patterns are explored through both theoretical analysis and practical implementation, emphasizing their relevance and applicability to real-world software development.
In alignment with the course’s objectives, several learning outcomes are directly associated with the conceptualization and application of design patterns. These outcomes, outlined in the official course syllabus, are summarized in Table~\ref{tab:learning-outcomes-patterns}. While the course encompasses six learning outcomes in total, only three of them are directly addressed and assessed during the first instructional unit focused on design patterns.

\begin{table*}[ht]
\centering
\scriptsize
\caption{Learning Outcomes Related to Design Patterns}
\begin{tabularx}{\textwidth}{@{}lX@{}}
\toprule
\textbf{Learning Outcome} & \textbf{Description} \\
\midrule
LO1 & Apply design patterns to create software systems that meet performance and security standards. \\

LO2 & Identify potential design patterns within already implemented software solutions. \\

LO3 & Implement a set of algorithms based on given problem specifications. \\

LO4 & Optimize existing algorithms to achieve optimal performance according to problem constraints. \\

LO5 & Define a comprehensive set of tests for complex algorithms based on given requirements. \\

LO6 & Characterize design patterns based on their applicability to different types of problems. \\
\bottomrule
\end{tabularx}
\label{tab:learning-outcomes-patterns}
\end{table*}

The course adopts an active learning methodology, combining lectures, labs, and project-based work. Students are assessed through a mix of written exams, collaborative projects, and presentations. The learning process is supported by key bibliographic resources, including the seminal GoF book by Gamma et al., along with complementary texts by Freeman and Robson, and Fowler. This approach fosters both technical competence and critical thinking in the use of design patterns.

\subsection{Characterization of the Student Group}

To better understand the composition and prior experience of the student group, a diagnostic instrument was administered at the beginning of the course. This \textit{characterization form} was completed by 18 out of 20 enrolled students and was designed to gather information on their academic background, programming experience, familiarity with design patterns, and perceived difficulties in the subject. The responses provide important context for interpreting the implementation and effectiveness of the Aronson cooperative learning methodology.

Among the respondents, 13 students were enrolled in the \textit{Ingeniería Civil en Computación e Informática} (ICCI) program and 5 in the \textit{Ingeniería en Tecnologías de la Información} (ITI) program. Most students were in their fifth semester, although a few were in the fourth, sixth, or eighth. All participants reported having taken previous programming courses, with self-reported software development experience ranging from none to approximately three years. The most commonly used programming languages included Java, Python, and C++, with some students also indicating experience with JavaScript, Unity/C\#, and GDScript.

While all students had at least heard of GoF (Gang of Four) design patterns prior to the course, the depth of their knowledge and practical exposure varied considerably. The most frequently recognized patterns were \textit{Singleton}, \textit{Factory Method}, \textit{Observer}, \textit{Strategy}, and \textit{Visitor}. A few students had implemented these patterns in academic or personal projects, but the majority lacked hands-on experience. Behavioral and creational patterns were more widely known and applied, while structural and compound patterns, such as \textit{Composite}, \textit{Proxy}, and \textit{Chain of Responsibility}, were less familiar. Many students expressed uncertainty about when and how to use patterns in concrete software development scenarios.

The characterization data also shed light on common learning difficulties. Several students pointed to the high cognitive load involved in memorizing patterns, especially when trying to differentiate between those with overlapping purposes or similar structural elements. Others reported that limited prior practice and a lack of clarity in the distinction between design principles and design patterns hindered their understanding. Additional challenges included applying patterns in UML diagrams and integrating them within larger architectural contexts.

\subsection{Pedagogical Rationale: A Professor’s Perspective on the Aronson Technique for Teaching Design Patterns}

To illustrate the motivation and reasoning behind the adoption of the Aronson cooperative learning technique, this study introduces a fictional character: \textbf{Professor Margareth}, a committed and reflective software engineering instructor. Through her narrative, we aim to present a realistic perspective grounded in actual classroom challenges and pedagogical decisions. By situating the teaching case in her voice, we contextualize the rationale for selecting cooperative learning as a strategy to address common difficulties students face when learning design patterns.

Before introducing the method itself, Professor Margareth administered a short \textit{characterization form} to her students. This 10-minute diagnostic activity aimed to gather baseline information on their prior programming experience and familiarity with design patterns. The responses helped her refine the group configurations and adjust her instructional approach based on the diversity of background knowledge within the classroom.

In the first instructional session, Professor Margareth formally introduced the Aronson technique using a concise 15-minute PowerPoint presentation. She explained its core structure: students would be divided into small base groups, with each member becoming an \say{expert} on a portion of the course material, then returning to teach it to their peers. The class responded with interest and curiosity to the notion of rotating between learner and instructor roles.

Margareth had long observed that students struggled to grasp design patterns. Many found them abstract or disconnected from practical development experience. Confusion often arose between the structure and purpose of individual patterns, or between patterns and broader design principles. Traditional teaching methods offered limited opportunities for students to articulate ideas in their own words or engage meaningfully with peers during class.

The Aronson technique, she believed, directly addressed these concerns. It promotes individual accountability, as each student must learn their assigned topic well enough to teach it. It also reinforces understanding through repetition and peer explanation, particularly valuable for internalizing nuanced distinctions between patterns such as \textit{Strategy}, \textit{Observer}, and \textit{Visitor}. The cooperative format supports diverse learning paces and strengthens conceptual clarity by fostering discussion, comparison, and reflection.

From a pedagogical standpoint, Margareth viewed the technique as well-suited to a topic requiring both theoretical understanding and practical application. The expert-group phase encouraged focused study of individual patterns, while the base-group phase ensured synthesis and knowledge transfer. This structure mirrors the dual nature of design patterns themselves: comprehending their intent and applying them in real-world contexts.

\subsubsection{Session 1: Creational Design Patterns}

The first content-focused session using the Aronson technique centered on Creational design patterns. Professor Margareth selected four foundational patterns for this activity: \textit{Abstract Factory}, \textit{Factory Method}, \textit{Singleton}, and \textit{Builder}. These patterns were carefully chosen not only for their prevalence in professional software development but also because they tend to be confusing for students due to their conceptual proximity.

Margareth justified the selection of these patterns on pedagogical grounds. \textit{Singleton} and \textit{Factory Method} are commonly introduced early in design pattern courses due to their relative simplicity and frequency of use. However, students often struggle to distinguish between them, especially regarding instantiation control and interface abstraction. To confront this, she provided contrasting use-case scenarios and UML diagrams extracted from the slides and guides shared in the session. 

\textit{Abstract Factory} and \textit{Builder} were introduced to expand students’ understanding of object creation in more complex systems. These patterns help illuminate concepts like product families and step-by-step construction, which are not always intuitive for learners accustomed to procedural or straightforward object-oriented code. The examples used in the provided material included illustrations such as factories producing cross-platform UI elements and builders assembling multi-step textual reports, both designed to ground abstract ideas in realistic domains.

To scaffold the expert group work, Margareth used curated slides from the shared guide, emphasizing key components: pattern intent, participants, class and sequence diagrams, and at least one complete code example. The examples were in Java, aligning with the students’ programming background. Each expert group received a printed and digital version of their assigned pattern’s material, with guiding questions such as:

\begin{itemize}
    \item What problem does this pattern solve?
    \item What are the main components and their roles?
    \item How does this pattern differ from the others studied in this session?
    \item Can you identify a scenario in which this pattern is particularly useful?
\end{itemize}

This design enabled students to prepare effectively before teaching their peers in the base groups. Margareth noted that the presence of real code and domain-relevant analogies helped students connect the pattern structure to practical applications—particularly helpful for those with less prior exposure to design patterns. Her pedagogical aim was not just recognition of the pattern name and structure, but deeper conceptual understanding through active reconstruction of knowledge.

\subsubsection{Session 2: Structural Design Patterns}

In the second week of the Aronson intervention, Professor Margareth focused on the family of \textit{structural design patterns}, introducing the patterns \textit{Decorator}, \textit{Facade}, \textit{Adapter}, and \textit{Proxy}. These patterns help organize classes and objects into larger structures while promoting flexibility and reusability. From previous semesters, Margareth had observed that students often struggled when applying structural patterns, particularly because many of them seemed interchangeable at first glance.

To address this, each base group received material and code examples specific to one structural pattern, accompanied by concise guides highlighting their \textit{intent}, \textit{structure}, \textit{applicability}, and \textit{collaborators}. The primary learning objective was not merely to memorize the patterns, but to develop the ability to recognize when each should be applied in real-world scenarios.

\vspace{0.2cm}
\noindent
Professor Margareth carefully selected problem descriptions that implicitly pointed to specific patterns. This method was intended to help students associate contextual cues with appropriate design solutions:
\begin{itemize}
    \item Situations involving \textit{adding multiple characteristics or responsibilities dynamically} hinted at the use of the \textbf{Decorator} pattern.
    \item Scenarios referring to \textit{a complex or legacy system that needed to be simplified} suggested the \textbf{Facade} pattern.
    \item Tasks highlighting \textit{incompatibility between classes or interfaces} pointed to the use of an \textbf{Adapter}.
    \item Descriptions involving \textit{access control, lazy initialization, or remote proxies} motivated the use of the \textbf{Proxy} pattern.
\end{itemize}

These examples encouraged analytical thinking and pattern identification based on intent rather than structure. During the expert group phase, students shared their interpretations, examined overlaps, and clarified distinctions, especially between patterns like \textit{Adapter} and \textit{Facade}, which can both hide complexity but serve different goals.

Upon returning to their base groups, students explained their assigned pattern and discussed the case problems collaboratively. The peer-teaching process allowed misconceptions to surface and be corrected, while also promoting metacognitive reflection on each student’s learning.

Professor Margareth noted an improvement in confidence and precision during group discussions compared to the previous week. She attributed this not only to increased familiarity with the method, but also to the contextual anchoring of the examples. Although challenges remained, especially in differentiating patterns with similar diagrams, the approach succeeded in making the patterns more accessible and relevant.

\subsubsection{Session 3: Behavioral Design Patterns}

In the third session, Professor Margareth introduced behavioral design patterns, known for managing algorithms, responsibilities, and object communication. These patterns are conceptually deeper and often more abstract, making them particularly challenging for students to grasp and implement effectively.

Four patterns were selected for this phase: \textit{Strategy}, \textit{Observer}, \textit{Template Method}, and \textit{Visitor}. Each was supported by structured guides and slides derived from canonical GoF resources, carefully adapted to the level of the students. Despite this scaffolding, behavioral patterns presented the most significant learning difficulties in the course.

\textit{Visitor} proved especially challenging. Its reliance on double dispatch clashed with the single dispatch nature of Java, used throughout the course, leading to confusion in implementing proper delegation. Even with simplified diagrams and walkthroughs, many students struggled to internalize the traversal logic and visitor method structure.

Similarly, \textit{Template Method} raised issues related to students’ weak foundations in object-oriented programming (OOP), especially inheritance and method overriding. Students often failed to see the benefit of controlling algorithm skeletons via abstract classes and had trouble distinguishing this from regular inheritance hierarchies.

\textit{Strategy}, while seemingly more accessible, often led to confusion with \textit{Decorator}. Both patterns rely on delegation and runtime composition, and the students frequently conflated the intent of encapsulating interchangeable behavior (Strategy) with dynamic interface enhancement (Decorator).

To mitigate these difficulties, Margareth emphasized comparative analysis during expert group discussions and provided guiding questions such as: \say{How does this pattern control flow or behavior?”, “What design principle is being applied?}, and \say{Could this be achieved using another structural or behavioral pattern, and what would be the trade-offs?}

Overall, this session solidified the importance of peer teaching inherent in the Aronson technique, as students with stronger OOP backgrounds became key supports within their base groups. This peer-led explanation, combined with incremental exposure to complexity, aligned well with the pedagogical aim of building durable understanding through collaboration and repetition.

\subsection{Formative Assessment Mechanisms: Mental Maps, Socrative Quizzes, and Self, Peer Evaluation}

To reinforce the cooperative learning process and gather evidence of both individual understanding and collective synthesis of design patterns, three complementary formative assessment strategies were implemented: group-created mental maps, individual multiple-choice quizzes administered via Socrative, and a self, peer evaluation instrument.

\paragraph{Group Mental Maps and Oral Presentations.} At the end of each weekly Aronson based learning cycle, students were tasked with collaboratively developing a mental map summarizing the design patterns studied during that session. Each base group had one week to prepare and present their map in class. These maps were expected to visually organize key aspects such as pattern intent, structure, participants, application scenarios, and relevant real-world examples. During the oral presentation, each group explained their diagram and justified their design decisions.

The evaluation of these presentations was guided by a checklist based rubric encompassing criteria such as conceptual accuracy, graphical organization, relevance of examples, depth of explanation, and overall creativity. This activity aimed to foster collaborative knowledge synthesis and to develop students’ ability to communicate complex design ideas visually and verbally.

\paragraph{Socrative Quizzes: Individual Assessment.} At the conclusion of each weekly module, students completed a 20 minute individual quiz using the Socrative platform. Each quiz consisted of multiple choice questions aligned with the design patterns addressed during the week. These assessments evaluated the students' ability to distinguish between patterns, apply them in context, and recognize their structural and functional differences.

Below are two representative questions from the Structural Patterns quiz:

\begin{itemize}
    \item \textbf{Question 1:} Which structural pattern allows adding new responsibilities to an object dynamically, without modifying its class?
    \begin{enumerate}
        \item Adapter
        \item Proxy
        \item Facade
        \item \textbf{Decorator (Correct answer)}
    \end{enumerate}

    \item \textbf{Question 2:} In a system where multiple incompatible interfaces must work together, which pattern is most appropriate?
    \begin{enumerate}
        \item Facade
        \item Singleton
        \item \textbf{Adapter (Correct answer)}
        \item Builder
    \end{enumerate}
\end{itemize}

The questions were closely aligned with those explored during the expert group discussions and served to assess individual comprehension, conceptual clarity, and reasoning ability.

\paragraph{Self and Peer Evaluation.} At the end of the Aronson based intervention, students completed a self and peer evaluation form designed to promote reflection and assess group collaboration. The instrument included twelve criteria grouped into five dimensions: responsibility, participation, collaboration, communication, and shared learning. Each item was rated using a four point Likert scale ranging from “Never” to “Always.”

This evaluation mechanism served several pedagogical functions. It encouraged metacognitive reflection, allowing students to assess their own level of engagement and contribution. It also supported transparency and accountability in group dynamics by surfacing perceptions of individual effort and peer collaboration. Importantly, it provided the instructor with a diagnostic view of group functioning, identifying highly engaged students, those requiring additional support, and potential imbalances within teams.

Within the cooperative framework of the Aronson method, where each learner is responsible for both acquiring and disseminating knowledge, the use of peer and self assessment reinforced individual responsibility and collective learning goals.

These three complementary strategies, collaborative synthesis, individual testing, and reflective evaluation, offered a robust and nuanced view of the learning process. Together, they ensured a balanced assessment of both content mastery and the soft skills essential for cooperative learning.

With these assessment mechanisms in place, it became possible to systematically identify common conceptual and practical difficulties faced by students in learning design patterns. These issues emerged across multiple evaluative contexts: in quiz responses, group discussions, concept map presentations, and reflective self assessments. To better visualize these findings, Table~\ref{tab:pattern-difficulties} presents a summary of the most frequent learning obstacles observed, organized by pattern type: creational, structural, and behavioral. This synthesis forms the basis for the more in-depth evaluative analysis that follows.

\begin{table*}[!ht]
\centering
\scriptsize
\caption{Summary of Student Difficulties with Design Patterns}
\begin{tabularx}{\textwidth}{>{\raggedright\arraybackslash}p{4cm} X}
\toprule
\textbf{Pattern Type} & \textbf{Common Student Difficulties} \\
\midrule
\textbf{Creational:} \newline \textit{Singleton, Factory Method} & 
\begin{itemize}
    \item Confusion between instantiation control and interface abstraction.
    \item Difficulty understanding when to use \textit{Factory Method} versus \textit{Abstract Factory}.
    \item Misinterpretation of pattern responsibilities due to lack of practical examples.
\end{itemize} \\

\textbf{Creational:} \newline \textit{Abstract Factory, Builder} &
\begin{itemize}
    \item Struggles with understanding product families and step-by-step construction.
    \item Difficulty in visualizing the benefit of these patterns over simple object creation.
\end{itemize} \\

\textbf{Structural:} \newline \textit{Adapter, Decorator, Proxy, Facade} &
\begin{itemize}
    \item Overlapping intent caused confusion (e.g., \textit{Adapter} vs. \textit{Facade}).
    \item Lack of clarity on pattern motivation from problem statements.
    \item Difficulty selecting the appropriate pattern from similar options.
\end{itemize} \\

\textbf{Behavioral:} \newline \textit{Visitor} &
\begin{itemize}
    \item Major issues with double dispatch implementation in Java.
    \item Confusion with method delegation and class traversal.
\end{itemize} \\

\textbf{Behavioral:} \newline \textit{Template Method} &
\begin{itemize}
    \item Weak understanding of OOP principles, especially inheritance and method overriding.
    \item Difficulty distinguishing from general inheritance structures.
\end{itemize} \\

\textbf{Behavioral:} \newline \textit{Strategy} &
\begin{itemize}
    \item Frequent confusion with \textit{Decorator} due to similar use of composition.
    \item Misunderstanding of encapsulated behavior vs. dynamic enhancement.
\end{itemize} \\
\bottomrule
\end{tabularx}
\label{tab:pattern-difficulties}
\end{table*}

The evaluation strategies implemented throughout the Aronson based intervention were designed not only to assess student achievement, but also to reveal the specific conceptual challenges encountered during the learning process. Analyzing the outcomes of Socrative quizzes, mental maps, and peer, self assessments offered a multifaceted perspective on student engagement and comprehension.

The Socrative quizzes, administered weekly, tested both recognition and contextual application of design patterns. Questions asked students to identify the most appropriate pattern for given software scenarios and to compare structural and functional differences among similar patterns. The results highlighted persistent confusion around patterns such as \textit{Singleton} and \textit{Factory Method}, where students struggled to differentiate between instantiation control and interface abstraction. Similarly, errors in distinguishing between \textit{Adapter} and \textit{Facade} revealed overlapping understandings of structural intent, often based on superficial familiarity rather than deeper comprehension. These difficulties echoed the patterns of misunderstanding summarized in Table~\ref{tab:pattern-difficulties}, confirming that abstract intent, rather than implementation syntax, was the main obstacle.

In contrast, mental map presentations emphasized collaborative synthesis and conceptual articulation. After one week of preparation, students presented diagrams connecting key ideas and structures of the assigned patterns. While some maps demonstrated strong comprehension, others, particularly those covering \textit{Builder} and \textit{Abstract Factory}, lacked clarity in representing product families or step by step construction. Maps on structural patterns often misrepresented participant roles or conflated multiple pattern intents. Notably, maps covering \textit{Visitor} frequently failed to capture correct delegation logic or traversal flows, revealing confusion with double dispatch, a known challenge in Java that significantly hindered understanding.

Self and peer evaluations provided insight into students’ metacognitive and social engagement. Students tasked with complex patterns like \textit{Visitor} or \textit{Template Method} often reported low confidence in their ability to teach peers, despite evident effort. This reflected a lack of foundational understanding in key object oriented principles such as delegation and inheritance. Conversely, students assigned to more familiar patterns like \textit{Singleton} or \textit{Strategy} expressed greater confidence, though sometimes this resulted in an overestimation of their actual understanding.

Cross referencing the outcomes from these evaluations confirmed the utility of the Aronson technique not only for promoting active and cooperative learning, but also for surfacing deep rooted misconceptions that might remain hidden in traditional instruction. Socrative quizzes provided immediate diagnostic feedback, mental maps revealed students' structural and conceptual connections, and the reflective evaluations surfaced emotional and cognitive responses to the learning process.

Collectively, these instruments provided Professor Margareth with a detailed picture of how students engaged with design patterns, highlighting both achievements and persistent gaps. They reinforced the value of using diverse formative tools in conjunction with cooperative learning to both foster understanding and inform pedagogical refinement.

\section{Teaching Case B: OSI Model in Networks with Aronson} \label{caseB}
This section presents the implementation of Aronson's cooperative learning technique in the Computer Networks course, focusing on its pedagogical foundation, instructional design, and preliminary impact. The context, course structure, and student group characteristics are detailed to provide a comprehensive basis for evaluating the effectiveness of this teaching strategy.

\subsection{Course Description and Context}

The Computer Networks course is a core subject in the eighth semester curriculum for students in the \textit{Ingeniería Civil en Computación e Informática} (ICCI) and \textit{Ingeniería en Tecnologías de la Información} (ITI) programs, as it is part of the curriculum related to the area of IT platforms. The course consists of 5 SCT (Transferable Credit System) credits, a Chilean standard for quantifying student workload based on the time required to achieve specific learning outcomes in higher education. The course is designed to develop students' skills in the design and implementation of LAN (local area network) and WAN (wide area network) communication networks, with special emphasis on the use of network architecture design principles and industry best practices to ensure efficient and secure network systems.

A fundamental pillar for understanding computer networks is the study of the OSI (Open Systems Interconnection) Model, which groups network functionalities into seven distinct layers: physical, data link, network, transport, session, presentation, and application. Students learn to understand and apply the principles of each layer to design, implement, and troubleshoot scalable and maintainable network infrastructures. These layers are explored through theoretical analysis and practical implementation, highlighting their relevance and applicability in configuring and managing networks in real-world situations.

Regarding the objectives of a networking course, several learning outcomes are directly related to the conceptualization and application of the OSI Model layers. These outcomes, essential for a complete understanding of network architecture, are summarized below (see Table~\ref{tab:learning-outcomes-nc}). While the course covers these eight learning outcomes in total, only one of them is directly addressed and assessed during the first learning unit focused on the OSI model, as they are typically the primary focus during the initial configuration phase of a basic network infrastructure.

\begin{table*}[ht]
\centering
\scriptsize
\caption{Learning Outcomes Related to Networks Computer}
\begin{tabularx}{\textwidth}{@{}lX@{}}
\toprule
\textbf{Learning Outcome} & \textbf{Description} \\
\midrule
LO1 & Understand basic networking concepts and technologies. \\

LO2 & Configure and implement LAN switches and wireless technologies using appropriate protocols and standards. \\

LO3 & Configure routers and troubleshoot routing issues. \\

LO4 & Implement wide area network (WAN) technologies on medium-sized networks, including security, troubleshooting processes, and protocols. \\

LO5 & Propose a network infrastructure with sufficient detail for infrastructure architects to complete. \\

LO6 & Identify the objectives and requirements of ICT solutions. \\

LO7 & Select appropriate processes, techniques, and tools according to requirements. \\

LO8 & Develop the most suitable technological solution based on the problem characteristics and available resources. \\
\bottomrule
\end{tabularx}
\label{tab:learning-outcomes-nc}
\end{table*}

\subsection{Characterization of the Student Group}

The study group was characterized using a \textit{characterization form}, initially distributed before the first session using the Aronson method. The objective of administering this form was to better understand the composition and prior experience of the student group. The form was completed by all 15 students enrolled in the course and was designed to gather information about their academic background, experience with computer networks, and familiarity with general IT infrastructure concepts. Student responses provided important context for interpreting the implementation and effectiveness of Aronson's cooperative learning methodology.

The characterization data were distributed among 12 students from \textit{Ingeniería Civil en Computación e Informática} (ICCI) and three from \textit{Ingeniería en Tecnologías de la Información} (ITI) program. Most of the students were in their eighth semester, and some were in their sixth or seventh semester. All participants had previously taken courses in computer architecture and operating systems, and their experience with computer networking concepts ranged from 0 to 1 year.

The characterization also showed that all students had heard about the basic concepts of the OSI model in the course, but their level of knowledge and experience varied considerably. The most frequently mentioned concepts were IP (Internet Protocol) addresses, HTTPS (HyperText Transfer Protocol Secure), MAC (Media Access Control) addresses, and hosts. Some students had implemented these concepts in courses from other semesters of their program, such as software development, where virtual containers like Docker\footnote{For more information, visit: \url{https://www.docker.com}} are configured to host applications based on microservices architectures. A significant percentage of students expressed uncertainty about computer networking concepts and when and how to apply them in practical contexts.

\subsection{Pedagogical Rationale: A Professor’s Perspective on the Aronson Technique for Teaching Computer Network}

To motivate the use of Aronson's cooperative learning technique, this study presents Professor Paul, a computer science instructor specializing in networking systems. His account illustrates the pedagogical reasoning behind applying cooperative learning to address students' difficulties with the OSI model in a computer networking course.

Before beginning the instructional session, Paul conducted a brief 10-minute assessment to characterize the students' prior knowledge. This activity was completed by all 15 students participating in the course. Using the data obtained, Paul identified knowledge gaps and formed balanced groups, ensuring a diversity of experience levels in each group. This approach sought to promote peer support and equitable participation, reinforcing his belief that cooperative learning would improve participation and deepen students' understanding of the OSI model.

Then, in the first instructional session, Professor Paul formally introduced the Aronson cooperative learning technique using a concise 15-minute PowerPoint presentation. He explained its core structure: students would be divided into small base groups, with each student becoming an expert on one of the OSI model layers. After studying their assigned layer in depth—such as the physical, data link, or transport layer—they would return to their base group to teach it to their peers. The class responded with curiosity and engagement, intrigued by the idea of rotating between learner and instructor roles.

Paul had observed from previous semesters that students often struggled to conceptually understand the OSI model. Many found layered abstraction difficult to internalize, especially when distinguishing between similar functionalities of protocols in adjacent layers. The gap between theoretical models and practical networking assignments frequently led to confusion or a superficial understanding. Traditional lectures rarely provided students with the opportunity to express their ideas in their own words or engage in meaningful peer discussions. According to Paul, that Aronson's technique should directly address these problems by promoting individual accountability (each student had to understand the assigned topic well enough to teach it) and reinforcing learning through repetition and peer explanation. This is especially valuable for clarifying subtle distinctions between layered protocols, such as between the network layer and the transport layer, or between the data link layer and the physical layer.

\subsubsection{Session 1: Application, Presentation, and Session Layers of the OSI Model}

The first content-focused session using the Aronson cooperative learning technique centered on the upper layers of the OSI model: \textbf{Application}, \textbf{Presentation}, and \textbf{Session}. Professor Paul selected these layers as the initial focus of cooperative learning because, although they are conceptually closer to the types of applications students interact with daily, they often generate confusion due to their abstract and overlapping roles in network communication.

In this session, each of the five expert group students was assigned a specific protocol related to these layers: \textbf{HTTPS} (Hypertext Transfer Protocol Secure), \textbf{SSH} (Secure Shell), \textbf{SSL} (Secure Sockets Layer), \textbf{RPC} (Remote Procedure Call), and \textbf{FTP} (File Transfer Protocol). These protocols were chosen for their relevance in real-world applications and their connection to critical functions such as secure communication, remote access, file transfer, and distributed computing.

Paul justified the selection based on both pedagogical value and conceptual challenge. For instance, while students may have encountered \textbf{HTTPS} and \textbf{FTP} during web or file access, few had reflected on their underlying mechanisms or how they relate to OSI functionalities. Similarly, concepts such as encrypted shell sessions with \textbf{SSH}, transport layer security through \textbf{SSL}, or inter-process communication via \textbf{RPC} were either unfamiliar or misunderstood, despite their foundational role in distributed systems.

To support understanding, each expert received structured learning materials that included concise explanations, protocol diagrams, and real-world analogies—such as \textbf{HTTPS} in secure web browsing, \textbf{SSH} for remote terminal access, or \textbf{RPC} for distributed function calls. Materials were complemented by guiding questions, such as:

\begin{itemize}
    \item What problem does this protocol solve?
    \item What OSI layer(s) does it operate at, and why?
    \item How does it ensure secure, reliable, or efficient communication?
    \item In what real-world scenarios is this protocol commonly used?
\end{itemize}

This structure enabled students to deeply explore their assigned protocol before returning to their base group to teach it. Paul noted that distributing the protocols across group students not only encouraged accountability, but also fostered meaningful peer-to-peer exchanges. His pedagogical aim was not only for students to recognize protocol names or uses, but to construct a more integrated understanding of how the upper OSI layers enable human-centered, secure, and structured communication over networks.

\subsubsection{Session 2: Transport and Network Layers of the OSI Model}

The second content-focused session using the Aronson cooperative learning technique addressed the intermediate layers of the OSI model: \textbf{Transport} and \textbf{Network}. Professor Paul selected these layers because they are critical for enabling reliable communication across networks, yet often pose challenges for students due to the complexity of protocol interactions and routing logic.

This session focused on five key data transmission protocols, each assigned to an expert group member: \textbf{TCP} (Transmission Control Protocol), \textbf{UDP} (User Datagram Protocol), \textbf{IP/ICMP} (Internet Protocol / Internet Control Message Protocol), \textbf{EIGRP} (Enhanced Interior Gateway Routing Protocol), and \textbf{OSPF} (Open Shortest Path First). These protocols were chosen to illustrate essential aspects of both data transport between applications and routing decisions within and across networks.

Expert groups began by analyzing the two main transport protocols. The \textbf{Transport Layer} was explored through a comparison of \textbf{TCP}, which offers reliable, connection-oriented data transfer with error checking and flow control, and \textbf{UDP}, which provides a faster, connectionless alternative with lower overhead. Practical use cases were discussed, such as file transfers over \textbf{TCP} and live video or audio streaming over \textbf{UDP}.

For the \textbf{Network Layer}, students studied how data is addressed, routed, and forwarded through networks using protocols like \textbf{IP} and \textbf{ICMP}. They also examined internal and external routing strategies by exploring \textbf{EIGRP}, an advanced distance-vector protocol used within organizations, and \textbf{OSPF}, a link-state routing protocol known for its efficiency in large-scale IP networks. The role of \textbf{ICMP} was clarified through examples of network diagnostics like \texttt{ping} and \texttt{traceroute}.

Paul provided learning materials tailored to each protocol, including annotated diagrams (e.g., TCP/UDP headers and routing tables), simplified routing topologies, and illustrative flowcharts. Each expert group received printed and digital guides accompanied by structured questions, such as:

\begin{itemize}
    \item What are the core responsibilities of this protocol within its layer?
    \item How does this protocol handle reliability, routing, or error reporting?
    \item How does it interact with other protocols across the OSI layers?
    \item In what types of network scenarios is this protocol especially useful?
\end{itemize}

This cooperative approach allowed students to explore both the vertical interactions between layers and the horizontal specialization of each protocol. By teaching their assigned protocol to their base groups, students developed not only technical knowledge but also the ability to synthesize and explain operational trade-offs. Paul observed that the peer-led discussions led to more nuanced questions about protocol performance, security implications, and real-world deployment contexts, demonstrating a deeper and more integrated understanding of the Transport and Network layers.

\subsubsection{Session 3: Data Link and Physical Layers of the OSI Model}

The third and final content-focused session using the Aronson cooperative learning technique addressed the lower layers of the OSI model: \textbf{Data Link} and \textbf{Physical}. Professor Paul selected this focus to help students understand how digital information is physically transmitted and how devices coordinate access to shared transmission media — aspects often overlooked in higher-level discussions of networking.

This session centered on five key technologies and protocols, each assigned to an expert group member: \textbf{Ethernet}, \textbf{CSMA/CD} (Carrier Sense Multiple Access with Collision Detection), \textbf{PPP} (Point-to-Point Protocol), \textbf{Token Ring}, and a combined study of \textbf{Physical Layer elements} including transmission media, hardware components, and signaling standards.

The expert group studying \textbf{Ethernet} examined how data is framed and transmitted in LAN environments using MAC addresses and standardized frame formats. The \textbf{CSMA/CD} group focused on the access method used in early Ethernet to avoid collisions in shared channels, discussing its relevance in historical and modern contexts. The \textbf{PPP} group explored how this protocol facilitates direct communication over point-to-point links, such as between routers and modems, with features like authentication and error detection.

The group assigned to \textbf{Token Ring} studied how access to the medium is controlled through a token-passing mechanism — a contrast to contention-based approaches like CSMA/CD. Though largely obsolete in practice, Token Ring was included to illustrate the evolution of media access control strategies. The final expert group addressed the core elements of the \textbf{Physical Layer}, including different types of transmission media (e.g., copper cables, fiber optics), physical connectors, signal modulation, and the standards (such as IEEE 802.3) that govern how bits are encoded and transmitted as electrical or optical signals.

Paul provided each expert group with visual aids such as signal diagrams, frame structure breakdowns, and real hardware photos. The learning guides included structured prompts to guide inquiry, such as:

\begin{itemize}
    \item What is the main function of this protocol or technology at its OSI layer?
    \item How does it manage or enable communication over physical media?
    \item What are its advantages or limitations compared to alternatives?
    \item How does this technology influence network design or performance?
\end{itemize}

Through this session, students engaged with the tangible, physical realities of computer networking. The hands-on examples and physical analogies such as comparing fiber optics to highways or MAC addresses to postal addresses helped demystify concepts typically perceived as technical or low-level. By teaching their peers, students demonstrated not only protocol comprehension but also an appreciation for the foundational role of the lower OSI layers in enabling reliable and efficient data communication.

\subsection{Formative Assessment Mechanisms: Mental Maps, Socrative Quizzes, and Self, Peer Evaluation}

To reinforce the cooperative learning process and gather evidence of both individual understanding and collective synthesis of networking concepts, three complementary formative assessment strategies were implemented: group-created mental maps, individual multiple-choice quizzes administered via Socrative, and a self and peer evaluation instrument.

\paragraph{Group Mental Maps and Oral Presentations.} 
At the end of each weekly Aronson-based learning cycle, students were tasked with collaboratively developing a mental map summarizing the key protocols and functions studied during that session. Each base group had one week to prepare and present their map in class. These maps were expected to visually organize concepts such as protocol purpose, OSI layer assignment, communication flow, use cases, and real-world applications (e.g., how \textbf{TCP} ensures reliable transmission, or how \textbf{Ethernet} handles data framing).

During the oral presentation, each group explained their map and justified its organization. The evaluation was guided by a rubric covering criteria such as conceptual accuracy, graphical clarity, relevance of examples, depth of explanation, and creativity. This activity aimed to foster collaborative synthesis and to develop students' ability to articulate complex networking ideas both visually and verbally.

\paragraph{Socrative Quizzes: Individual Assessment.}
At the conclusion of each weekly module, students completed a 20-minute individual quiz using the Socrative platform. Each quiz consisted of multiple-choice questions aligned with the protocols and layers studied that week. These assessments evaluated the student’s ability to identify protocol functions, assign them to the correct OSI layer, and apply their characteristics to practical networking scenarios.

Below are two representative questions from the Transport and Network Layers quiz:

\begin{itemize}
    \item \textbf{Question 1:} Which protocol ensures reliable, connection-oriented data transmission between applications?
    \begin{enumerate}
        \item \textbf{TCP (Correct answer)}
        \item UDP
        \item IP
        \item ICMP
    \end{enumerate}

    \item \textbf{Question 2:} What is the main function of the OSPF protocol?
    \begin{enumerate}
        \item Error detection at the data link layer
        \item Transporting files between hosts
        \item \textbf{Determining the best routing paths within an autonomous system (Correct answer)}
        \item Translating domain names into IP addresses
    \end{enumerate}
\end{itemize}

These quizzes assessed both recognition and contextual application, reinforcing the conceptual distinctions between protocols like \textbf{TCP} vs. \textbf{UDP}, or \textbf{OSPF} vs. \textbf{EIGRP}.

\paragraph{Self and Peer Evaluation.}

At the end of the Aronson-based intervention in computer networks, students completed a self and peer evaluation designed to foster reflection and assess group collaboration. The instrument included twelve items grouped into five key dimensions: \textit{responsibility}, \textit{participation}, \textit{collaboration}, \textit{communication}, and \textit{shared learning}. Each was rated on a four-point Likert scale ranging from ``Never'' to ``Always.''

This evaluation mechanism served multiple pedagogical purposes. It encouraged students to reflect on their individual contributions and group engagement, promoting both metacognitive awareness and peer accountability. For the instructor, it provided diagnostic insights into team functioning, revealing highly engaged students, imbalances in group dynamics, and opportunities for targeted support.

Within the cooperative structure of the Aronson method, where each student becomes an expert in a specific networking protocol or concept (such as \textbf{TCP}, \textbf{UDP}, \textbf{ICMP}, or \textbf{OSPF}), the use of peer and self-assessment reinforced individual responsibility while fostering collective learning outcomes.

These three complementary strategies—collaborative synthesis (mental maps), individual testing (quizzes), and reflective evaluation—provided a robust view of student learning. It allowed for the assessment of both technical mastery and interpersonal skills essential to teamwork in computer networking.

These formative mechanisms also revealed recurring conceptual difficulties related to network protocol understanding and OSI layer interactions. Patterns of misunderstanding appeared consistently in quiz responses, group presentations, and self-evaluations. To visualize these challenges, Table~\ref{tab:network-difficulties} summarizes the most frequent obstacles identified during the intervention, organized by OSI layer and protocol type. This synthesis informs the evaluative discussion that follows.

\begin{table*}[!ht]
\centering
\scriptsize
\caption{Summary of Student Difficulties with OSI Layers and Network Protocols}
\begin{tabularx}{\textwidth}{>{\raggedright\arraybackslash}p{4cm} X}
\toprule
\textbf{OSI Layer or Protocol Group} & \textbf{Common Student Difficulties} \\
\midrule
\textbf{Application Layer:} \newline \textit{HTTP, FTP, SMTP, SSH, RPC} & 
\begin{itemize}
    \item Confusion between the protocol and the application (e.g., mistaking FTP as a file manager rather than a protocol).
    \item Unclear about client-server roles and communication flow.
    \item Difficulty distinguishing between protocols that provide similar services (e.g., HTTP vs. HTTPS).
\end{itemize} \\

\textbf{Presentation Layer:} \newline \textit{TLS, SSL, JSON, Encoding/Compression} &
\begin{itemize}
    \item Misunderstanding the role of data formatting vs. encryption.
    \item Difficulty identifying when this layer is active in real-world tools.
    \item Limited grasp of format negotiation and transformation processes.
\end{itemize} \\

\textbf{Session Layer:} \newline \textit{Session Management, Dialog Control} &
\begin{itemize}
    \item Abstractness made it hard to visualize session lifecycle.
    \item Students confused session roles with transport-layer responsibilities.
    \item Unclear examples in modern applications reduced perceived relevance.
\end{itemize} \\

\textbf{Transport Layer:} \newline \textit{TCP, UDP} &
\begin{itemize}
    \item Confusion between reliability and speed trade-offs.
    \item Difficulty understanding port addressing and connection management.
    \item Struggled to apply protocol choice to concrete scenarios (e.g., video calls vs. file transfers).
\end{itemize} \\

\textbf{Network Layer:} \newline \textit{IP, ICMP, OSPF, EIGRP, BGP} &
\begin{itemize}
    \item Lack of clarity on IP addressing vs. routing protocol roles.
    \item Confusion between internal (IGP) and external (EGP) routing.
    \item Misinterpretation of ICMP as an application-layer tool.
\end{itemize} \\

\textbf{Data Link Layer:} \newline \textit{Ethernet, CSMA/CD, PPP, Token Ring} &
\begin{itemize}
    \item Difficulty understanding framing and MAC addressing.
    \item Token Ring seen as obsolete and hard to contextualize.
    \item CSMA/CD often confused with basic broadcast mechanisms.
\end{itemize} \\

\textbf{Physical Layer:} \newline \textit{Transmission Media, Signals, Hardware Standards} &
\begin{itemize}
    \item Challenges linking physical components to higher-layer processes.
    \item Misconceptions about analog vs. digital signaling.
    \item Lack of understanding of standards such as IEEE 802.3.
\end{itemize} \\
\bottomrule
\end{tabularx}
\label{tab:network-difficulties}
\end{table*}

The evaluation strategies implemented throughout the Aronson-based intervention in computer networks were designed not only to assess student achievement but also to uncover specific conceptual challenges encountered during the learning process. Analyzing the results of Socrative quizzes, mental maps, and self and peer assessments provided a comprehensive view of student engagement and understanding.

The weekly Socrative quizzes assessed both recognition and contextual application of networking concepts. Questions asked students to identify the appropriate protocol or technology for given network scenarios and to compare the functions and roles of similar protocols. The results revealed persistent confusion around topics such as the differences between TCP and UDP, or the roles of routing protocols like OSPF versus RIP. Similarly, misconceptions about OSI layers, particularly the distinctions between data link and network layers, highlighted gaps in foundational understanding. These difficulties are summarized in Table~\ref{tab:network-difficulties}, confirming that conceptual clarity, rather than memorization of standards, was the main challenge.

In contrast, mental map presentations focused on collaborative synthesis and conceptual articulation. After a week of preparation, students presented diagrams connecting key ideas such as protocol interactions, packet flow, and network architecture. While some groups demonstrated strong understanding, others showed gaps in mapping complex processes like TCP’s three-way handshake or the workings of VLANs and subnetting. Maps on routing protocols often misrepresented routing tables or conflated dynamic and static routing concepts. These inaccuracies revealed areas needing targeted instruction.

Self and peer evaluations gave insight into students’ metacognitive and social engagement. Those assigned to more complex topics such as network security or advanced routing reported lower confidence in explaining these concepts, despite active participation. Conversely, students working on foundational topics like Ethernet or IP addressing showed higher confidence, though sometimes overestimated their depth of understanding.

Cross-referencing outcomes from these evaluations confirmed the effectiveness of the Aronson method in promoting active and cooperative learning while revealing deep-rooted misconceptions often overlooked in traditional lectures. Socrative quizzes provided timely diagnostic feedback, mental maps revealed structural and conceptual connections, and reflective evaluations surfaced students’ cognitive and emotional responses to learning.

Together, these tools offered the instructor a detailed picture of how students engaged with computer networking concepts, highlighting both strengths and persistent challenges. They emphasize the importance of diverse formative assessments combined with cooperative learning to enhance comprehension and guide pedagogical improvements.

\section{Analyzing Student Performance: Statistical and Pedagogical Perspectives} \label{analysis}
This section examines the impact of the Aronson Jigsaw technique on student performance through both quantitative and pedagogical lenses. Using statistical evidence from two cohorts, one studying software design patterns and the other the OSI model, we evaluate correlations between collaborative engagement and individual achievement. Beyond the numerical analysis, we discuss how the method may promote deeper learning through cognitive and motivational processes that extend beyond the scope of performance metrics alone.

\subsection{Pedagogical Impact of Collaborative Learning Across Courses}

We evaluate the pedagogical effectiveness of the Aronson cooperative learning technique across two distinct instructional contexts: the teaching of software design patterns and the OSI model in computer networks. In both cases, the analysis draws on quantitative indicators that reflect individual understanding, group collaboration, and peer-evaluated engagement within the 2025-I cohort. A comparative analysis with the 2023 and 2024 cohorts is also conducted to assess whether the implementation of the Aronson method corresponds to improved academic performance.

Our analysis is guided by the following hypotheses:

\begin{itemize}
    \item $H_0$: There is no statistically significant relationship between students’ final performance and their engagement levels measured through collaborative activities.
    \item $H_1$: Higher engagement levels, as captured by the Collaborative Learning Index (CLI), are positively associated with improved individual performance in the final evaluation.
\end{itemize}

To examine these hypotheses, we employ both descriptive and inferential statistical methods. Within the 2025 cohort, we collect formative indicators such as weekly quiz scores, group mental map evaluations, and peer/self-assessments of participation. These indicators are combined to construct the CLI, which serves as a proxy for each student’s level of active engagement.

For the inter-cohort analysis, we restrict the comparison to final grade (FG) scores, as previous cohorts did not include comprehensive formative assessments. The Shapiro–Wilk test is applied to assess the normality of CLI and FG distributions. Based on these results, we use either Pearson or Spearman correlation to evaluate the relationship between collaborative engagement and individual performance. For comparisons of FG across cohorts, the Wilcoxon rank-sum test is employed due to its suitability for small samples and non-normal data distributions.

This unified approach provides a coherent analytical framework to evaluate the effectiveness of the Aronson method across both technical domains, while also ensuring methodological consistency in the statistical treatment of learning outcomes.

To quantitatively assess students’ engagement and learning in the context of the Aronson cooperative technique, this section introduces and justifies a composite indicator called the \textbf{Collaborative Learning Index} (CLI). The purpose of this index is to synthesize key dimensions of the formative evaluation process into a single metric that reflects both individual understanding and collaborative participation. This metric is later used in the statistical analysis to investigate its relationship with final course performance.

The CLI integrates three formative components collected during the intervention:

\begin{itemize}
    \item \textbf{ISC (Individual Socrative Score):} Captures the student’s individual understanding of the design patterns through weekly multiple-choice quizzes.
    \item \textbf{GMMS (Group Mental Map Score):} Reflects the student’s contribution to the collaborative synthesis of knowledge via group-created mental maps and oral presentations.
    \item \textbf{SPES (Self and Peer Evaluation Score):} Represents perceived participation and collaboration as assessed through reflective self and co-evaluations.
\end{itemize}

These three dimensions are combined into a weighted linear formula:

\[
\text{CLI} = \alpha \cdot \text{ISC} + \beta \cdot \text{GMMS} + \gamma \cdot \text{SPES}
\]

The weights reflect a revised pedagogical rationale based on observations from the 2025 cohort:

\begin{itemize}
    \item \( \alpha = 0.4 \): Prioritizes individual accountability and mastery of concepts, as reflected by students' quiz performance.
    \item \( \beta = 0.5 \): Emphasizes the collaborative synthesis of knowledge, in line with the centrality of group work in the Aronson technique.
    \item \( \gamma = 0.1 \): Assigns reduced weight to peer and self-evaluations due to potential inflation effects. Most students awarded high scores to themselves and peers, possibly influenced by social bonds or a lack of critical self-reflection, thus undermining the discriminative value of this metric.
\end{itemize}

All input scores were normalized on a 100-point scale to ensure comparability. The resulting CLI values serve as a proxy for each student's level of active engagement and are used in subsequent inferential analyses to test correlations with final exam performance. Tables~\ref{tab:iac-values} and~\ref{tab:cli-values-networks} present the computed CLI values for students in both instructional contexts.

\begin{table}[!ht]
\scriptsize
\centering
\caption{Collaborative Learning Index (CLI) Components and Final Exam Grades — Patterns Course, Semester 2025-I}
\label{tab:iac-values}
\begin{tabularx}{0.48\textwidth}{l *{5}{>{\centering\arraybackslash}X}}
\toprule
\textbf{Student} & \textbf{ISC} & \textbf{GMMS} & \textbf{SPES} & \textbf{CLI} & \textbf{Final Grade (FG)} \\
\midrule
S1  & 64  & 98  & 100 & 81 & 54.0 \\
S2  & 86  & 76  & 100 & 83 & 78.0 \\
S3  & 80  & 64  & 70  & 73 & 50.0 \\
S4  & 82  & 70  & 100 & 79 & 54.0 \\
S5  & 78  & 86  & 100 & 83 & 46.0 \\
S6  & 84  & 94  & 100 & 90 & 66.0 \\
S7  & 72  & 84  & 100 & 80 & 52.0 \\
S8  & 94  & 98  & 100 & 96 & 50.0 \\
S9  & 86  & 100 & 100 & 93 & 86.0 \\
S10 & 86  & 100 & 100 & 93 & 44.0 \\
S11 & 50  & 28  & 60  & 42 & 50.0 \\
S12 & 84  & 70  & 100 & 80 & 48.0 \\
S13 & 48  & 68  & 100 & 61 & 46.0 \\
S14 & 100 & 86  & 100 & 94 & 74.0 \\
S15 & 98  & 100 & 100 & 99 & 82.0 \\
S16 & 44  & 100 & 100 & 72 & 90.0 \\
S17 & 44  & 78  & 100 & 63 & 60.0 \\
S18 & 60  & 98  & 100 & 79 & 50.0 \\
S19 & 58  & 100 & 100 & 79 & 60.0 \\
S20 & 64  & 90  & 100 & 78 & 60.0 \\
\bottomrule
\end{tabularx}
\end{table}

    \begin{table}[!ht]
    \centering
    \scriptsize
    \caption{Collaborative Learning Index (CLI) Components and Final Exam Grades — Networks Course, Semester 2025-I}
    \label{tab:cli-values-networks}
    \begin{tabularx}{\linewidth}{l *{4}{>{\centering\arraybackslash}X} >{\raggedleft\arraybackslash}X}
    \toprule
    \textbf{Student} & \textbf{ISC} & \textbf{GMMS} & \textbf{SPES} & \textbf{CLI} & \textbf{FG} \\
    \midrule
    S1  & 66.7 & 64  & 62 & 64.4 & 60.1 \\
    S2  & 76.7 & 79  & 78 & 77.6 & 81.0 \\
    S3  & 63.3 & 60  & 66 & 63.2 & 64.3 \\
    S4  & 76.7 & 79  & 73 & 76.0 & 79.9 \\
    S5  & 76.7 & 79  & 72 & 75.3 & 76.5 \\
    S6  & 56.7 & 53  & 52 & 54.4 & 62.0 \\
    S7  & 66.7 & 62  & 64 & 64.7 & 66.2 \\
    S8  & 56.7 & 53  & 50 & 53.7 & 65.9 \\
    S9  & 63.3 & 62  & 64 & 63.3 & 66.5 \\
    S10 & 70.0 & 65  & 62 & 65.7 & 73.5 \\
    S11 & 67.0 & 63  & 58 & 62.7 & 40.0 \\
    S12 & 72.0 & 69  & 64 & 68.3 & 70.2 \\
    S13 & 61.0 & 58  & 60 & 59.7 & 65.4 \\
    S14 & 64.0 & 66  & 65 & 65.0 & 72.7 \\
    S15 & 70.0 & 67  & 63 & 66.7 & 95.0 \\
    \bottomrule
    \end{tabularx}
    \end{table}

\subsection{Statistical Analysis of Learning Outcomes}

To perform the statistical analysis for both instructional cases (Software Design Patterns and OSI Model in Networks), we used R software\footnote{For more information, visit: \url{https://www.r-project.org/}}. To assess normality, the Shapiro--Wilk test was applied to both the Collaborative Learning Index (CLI) and final grade (FG) distributions.

In the Design Patterns case, CLI values yielded $W = 0.90966$ with a p-value of $0.06278$, indicating approximate normality. FG values showed $W = 0.86322$ with a p-value of $0.008949$, suggesting significant deviation from normality. For the Networks case, CLI values had $W = 0.9368$ and $p = 0.2157$ (normal), while FG values were non-normal with $W = 0.8421$ and $p = 0.0116$. Due to these results, Spearman’s rank correlation—a non-parametric method—was used in both cases to assess the relationship between CLI and FG.

In the Design Patterns cohort, Spearman’s correlation yielded $\rho = 0.195$ and $p = 0.4111$, while in the Networks cohort it was $\rho = 0.204$ with $p = 0.388$. Both results indicate weak, non-significant associations between CLI and FG.

\begin{figure}[ht]
    \centering
    \includegraphics[width=1\linewidth]{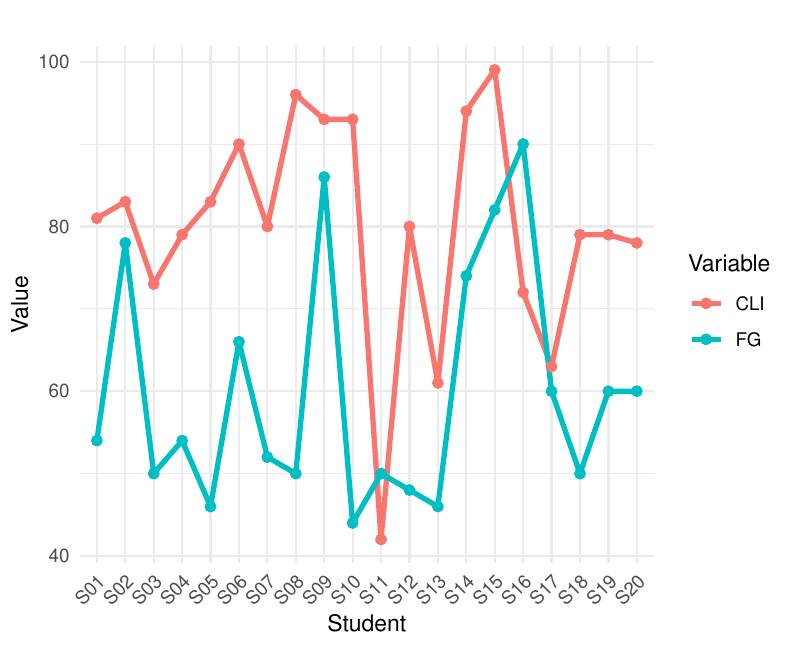}
    \caption{Correlation graph between CLI and FG across students}
    \label{fig:SPCorr}
\end{figure}

\begin{figure}[ht]
    \centering
    \includegraphics[width=1\linewidth]{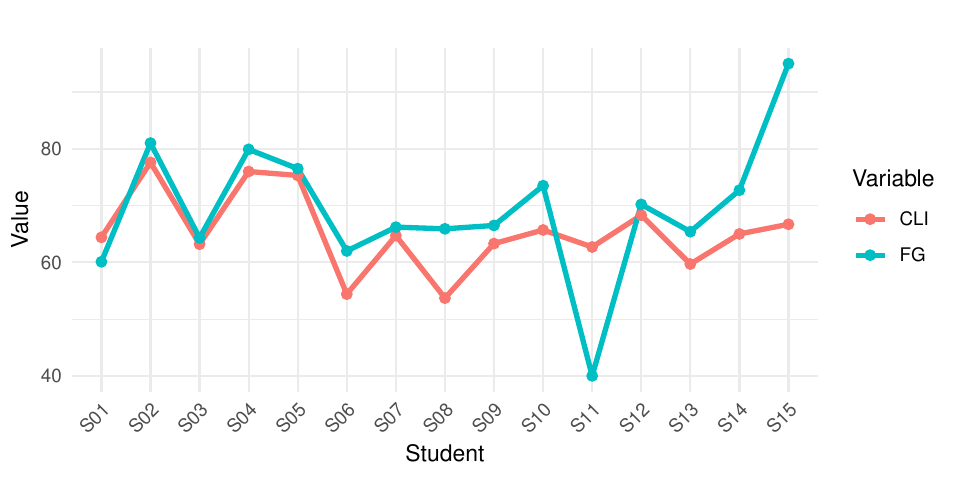}
    \caption{Correlation graph between CLI and FG across students}
    \label{fig:SPCorr-nc}
\end{figure}

Figure~\ref{fig:SPCorr} and Figure~\ref{fig:SPCorr-nc} present visualizations of CLI and FG values for both cohorts. Although curves display parallel fluctuation in some cases, the statistical results confirm a weak monotonic relationship.


Several trends support a nuanced interpretation. In the Design Patterns cohort, students like S6, S8, and S10 showed high CLI but low FG, aligning with \cite{Silva2020}, who noted that collaborative strategies may enhance engagement without guaranteeing deep understanding. In the Networks case, discrepancies such as S11's low FG (40.0) despite average CLI (62.7), and S15's high FG (95.0) with moderate CLI (66.7), suggest that individual mastery can diverge from group participation. Peer/self-evaluation scores (SPES) in both cohorts were clustered toward the upper end, potentially inflating the CLI and reducing its discriminative power, as previously reported by \cite{McMillan2021, Stonewall2024}.

Final evaluations placed emphasis on individual competencies, abstraction, design application, subnetting, and protocol analysis, that demand synthesis and problem-solving beyond what is typically reinforced in group-based activities. This underscores the need to complement collaborative methods with mechanisms that promote individual accountability \citep{Silva2020, Hwang2012}.

Despite the weak correlation between CLI and FG, the 2025-I Design Patterns cohort showed significantly higher final grades than previous cohorts. Wilcoxon rank-sum test comparisons revealed:

\begin{itemize}
    \item \textbf{2025-I vs 2024-II:} $p = 0.0187$
    \item \textbf{2025-I vs 2024-I:} $p = 0.0006$
    \item \textbf{2025-I vs 2023-II:} $p = 0.0095$
\end{itemize}

\begin{figure}[ht]
    \centering
    \includegraphics[width=1\linewidth]{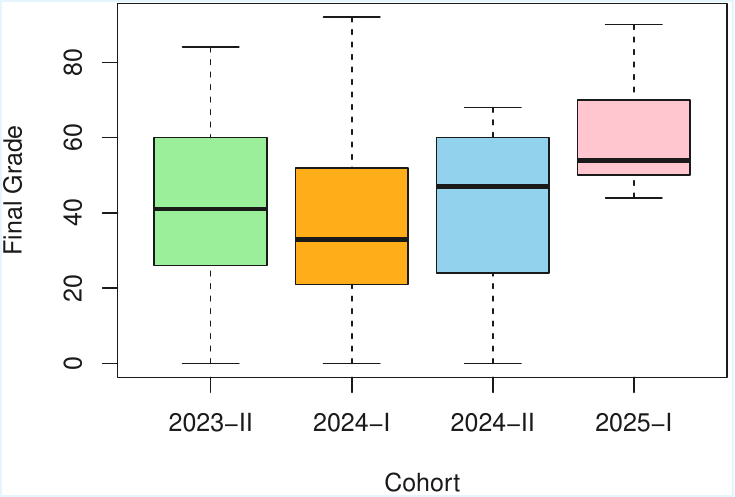}
    \caption{Boxplot of final grade distributions across cohorts}
    \label{fig:Boxplot1}
\end{figure}

Table~\ref{tab:grades-by-cohort} and Figure~\ref{fig:Boxplot1} illustrate this improvement, showing increased median FG and reduced dispersion in 2025-I. In contrast, the Networks cohort (2025-I) did not show statistically significant differences from previous years:

\begin{itemize}
    \item \textbf{2025-I vs 2024-II:} $p = 0.4505$
    \item \textbf{2025-I vs 2024-I:} $p = 0.216$
    \item \textbf{2025-I vs 2023-II:} $p = 0.2634$
\end{itemize}

\begin{figure}[ht]
    \centering
    \includegraphics[width=1\linewidth]{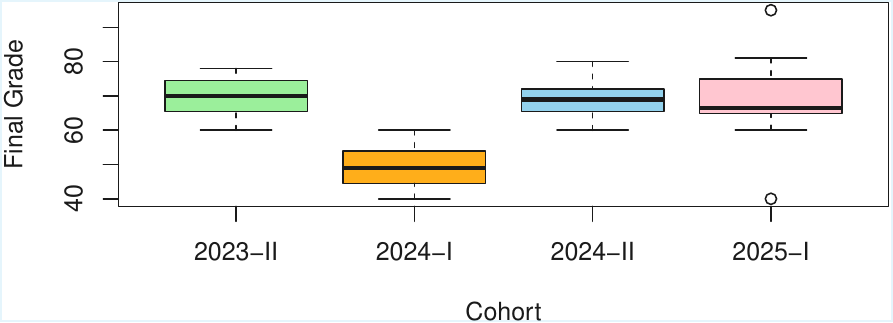}
    \caption{Boxplot of final grade distributions across cohorts}
    \label{fig:BoxPlot1-nc}
\end{figure}

However, the broader performance range observed in Figure~\ref{fig:BoxPlot1-nc} and detailed in Table~\ref{tab:grades-by-cohort2} suggests that the method enabled higher achievement for some students.

\begin{table}[!ht]
\centering
\scriptsize
\caption{Final grade distributions across cohorts (2023–2025) Patterns Course}
\begin{tabularx}{.45\textwidth}{X|X|X|X}
\toprule
\textbf{FG 2025-I} & \textbf{FG 2024-II} & \textbf{FG 2024-I} & \textbf{FG 2023-II} \\
\midrule
54  & 0   & 0   & 18 \\
78  & 66  & 34  & 28 \\
50  & 60  & 60  & 48 \\
54  & 56  & 14  & 68 \\
46  & 52  & 32  & 10 \\
66  & 46  & 38  & 46 \\
52  & 32  & 22  & 14 \\
50  & 42  & 92  & 26 \\
86  & 24  & 42  & 54 \\
44  & 24  & 32  & 28 \\
50  & 48  & 70  & 36 \\
48  & 46  & 88  & 76 \\
46  & 68  & 32  & 60 \\
74  & 0   & 20  & 0  \\
82  & 58  & 24  & 30 \\
90  & 62  & 16  & 54 \\
60  & 66  & 60  & 84 \\
50  & 0   & 36  & 60 \\
60  &     & 44  &    \\
60  &     & 0   &    \\
     &     & 70  &    \\
     &     & 8   &    \\
     &     & 52  &    \\
     &     & 48  &    \\
     &     & 22  &    \\
     &     & 0   &    \\
     &     & 30  &    \\
     &     & 8   &    \\
     &     & 22  &    \\
     &     & 0   &    \\
\bottomrule
\end{tabularx}
\label{tab:grades-by-cohort}
\end{table}
\begin{table}[!ht]
\centering
\scriptsize
\caption{Final grade distributions across cohorts (2023–2025) — Network Courses}
\begin{tabularx}{.48\textwidth}{X|X|X|X}
\toprule
\textbf{FG 2025-I (n=15)} & \textbf{FG 2024-II (n=18)} & \textbf{FG 2024-I (n=20)} & \textbf{FG 2023-II (n=20)} \\
\midrule
60.1 & 72  & 60  & 70 \\
81.0 & 80  & 58  & 78 \\
64.3 & 66  & 55  & 65 \\
79.9 & 76  & 48  & 77 \\
76.5 & 74  & 52  & 75 \\
62.0 & 60  & 42  & 64 \\
66.2 & 68  & 45  & 66 \\
65.9 & 64  & 40  & 68 \\
66.5 & 65  & 50  & 60 \\
73.5 & 69  & 46  & 70 \\
40.0 & 66  & 44  & 71 \\
70.2 & 72  & 55  & 76 \\
65.4 & 64  & 41  & 65 \\
72.7 & 70  & 49  & 68 \\
95.0 & 71  & 53  & 74 \\
      & 60  & 47  & 58 \\
      & 63  & 51  & 66 \\
      & 62  & 37  & 62 \\
      &     & 54  & 73 \\
      &     & 46  & 59 \\
\bottomrule
\end{tabularx}
\label{tab:grades-by-cohort2}
\end{table}

Although no statistically significant correlation was found between the Collaborative Learning Index (CLI) and final exam scores, this outcome underscores the complexity of assessing learning through summative metrics alone. The CLI reflects formative aspects of engagement, such as participation in quizzes, peer evaluations, and collaborative synthesis, which may not directly translate into final grades. 

Nevertheless, the use of the Aronson Jigsaw method appears to foster cognitive and motivational gains that contribute to deeper learning processes. This interpretation is consistent with previous findings reporting high levels of student satisfaction and perceived learning benefits when applying the Jigsaw technique in health education contexts \citep{LeyvaMoral2016}.

\subsection{Interpreting the Learning Impact}

The observed academic improvements in the 2025 cohorts, despite the weak statistical correlation between collaborative learning indicators and final exam performance, point to the influence of underlying cognitive, metacognitive, and affective processes. This section explores how the Aronson Jigsaw method may contribute to deeper learning outcomes through mechanisms not fully captured by quantitative metrics like the Collaborative Learning Index (CLI). Drawing on established educational psychology literature, we outline key factors that may explain the positive impact of this cooperative approach on individual student achievement.

\textit{Memory Consolidation through Distributed Retrieval Practice.} One of the key advantages of the Aronson method lies in its structure of peer teaching and distributed repetition. Each student must understand, explain, and reprocess the material from both the expert and base group perspectives. This process aligns with what cognitive psychology identifies as retrieval-based learning, a well-established mechanism for improving long-term memory retention \citep{Jeffrey2011}. In the context of programming concepts such as design patterns or OSI-layer functionalities, this distributed retrieval may strengthen conceptual frameworks that aid performance on final exams, even if these benefits are not directly observable in the CLI score.

\medskip

\noindent
\textit{Metacognitive Engagement and Teaching-as-Learning.} According to the \say{learning-by-teaching} paradigm \citep{Fiorella2023}, students who prepare to teach a topic engage in deeper cognitive processing. The Aronson technique, by design, places students in the role of \say{expert teachers}, which not only enhances accountability but also activates metacognitive skills such as planning, monitoring, and evaluating one’s own learning. This deeper level of processing helps students build more robust mental models that support transfer of knowledge to unfamiliar contexts, such as subnetting problems or design pattern identification in exam settings.

\medskip

\noindent
\textit{Affective and Motivational Gains.} Collaborative learning environments can also have affective benefits that indirectly improve performance. As \citet{Silva2020} note, group-based methods may enhance student motivation, engagement, and confidence, especially in challenging subjects such as computer programming and network systems. Increased motivation may result in greater effort during preparation for the final exam, independent of the immediate gains reflected in formative scores like the CLI.

\medskip

\noindent
\textit{Improved Self-Regulated Learning Habits.} Students immersed in collaborative methodologies often develop self-regulated learning strategies, including goal setting, resource management, and strategic reviewing. These habits, cultivated during the intervention and reinforced by the responsibility of teaching peers, may translate into improved preparation for final assessments, as students are better equipped to study independently and identify knowledge gaps.

\medskip

Despite the lack of a strong statistical correlation between the CLI and final exam performance in either course, students in the 2025 cohorts achieved significantly higher or more diverse final scores compared to previous semesters. This improvement may be attributed to the cognitive, metacognitive, and motivational benefits fostered by the Aronson Jigsaw method. The structured peer interaction and active engagement promoted by the technique likely contributed to more durable learning gains, even if not fully captured by short-term indicators like the CLI. These findings align with \citet{Silva2020}, who emphasize that cooperative learning environments enhance student persistence, conceptual integration, and long-term academic performance.

Nevertheless, the current study relies on two primary metrics: the \textit{Collaborative Learning Index} (CLI), capturing student contributions to group-based learning, and final exam grades, which reflect individual mastery. These indicators do not fully operationalize deeper constructs such as memory consolidation, metacognition, or affective engagement.

\section{Discussion} \label{discussion}
The application of the Aronson Jigsaw method in both software design patterns and computer networks courses revealed a nuanced relationship between collaborative engagement and individual performance. Although the Collaborative Learning Index (CLI) did not exhibit a statistically significant correlation with final exam scores in either case, students from the 2025 cohort attained higher final grades than those in previous academic periods. This suggests that the benefits of the intervention extend beyond what is captured by short-term indicators.

Several cognitive and motivational factors contribute to this outcome. The design of the Jigsaw method promotes repeated exposure to course material through peer instruction and reciprocal explanation. This distributed retrieval practice strengthens long-term memory by reinforcing conceptual understanding through active recall. As students alternate between learning from peers and teaching others, they engage with content at multiple cognitive levels, which supports deeper retention and improved application during final assessments.

The structure of the method also fosters metacognitive engagement. When students prepare to teach, they are prompted to plan, monitor, and evaluate their comprehension, which helps develop more robust internal frameworks. These metacognitive processes are not easily reflected in collaborative metrics like the CLI, but they may lead to stronger individual performance, especially in exams requiring conceptual transfer and problem-solving.

Beyond cognitive effects, the collaborative learning environment appears to offer motivational benefits. Increased engagement, confidence, and social cohesion may contribute to enhanced effort and persistence, particularly in courses perceived as difficult. These affective dimensions, although not directly observable in the CLI scores, likely played a role in students' exam preparation and final outcomes.


The analysis also revealed certain limitations. Self and peer evaluations tended to cluster at the higher end of the scale, reducing their ability to meaningfully distinguish between different levels of participation. Moreover, some students performed well in the final exam despite having low or moderate CLI scores. For example, in the networks course (see figure~\ref{fig:BoxPlot1-nc}, student S15 achieved a high final grade (95) with a relatively low CLI score (66). Similarly, in the design patterns course (see figure~\ref{fig:Boxplot1}), student S10 obtained one of the highest CLI values (91), reflecting consistent and active engagement in collaborative tasks, yet scored poorly in the final exam (58). These contrasting cases suggest that some students may compensate for limited collaborative engagement through strong prior knowledge or individual study skills, while others may actively participate in group activities without achieving deep individual understanding. Such patterns imply that collaborative participation does not benefit all students equally and that its impact may depend on individual differences as well as the alignment between group activities and the cognitive demands of summative assessments.

Final exams in both courses emphasized individual competencies such as abstraction, problem-solving, subnetting, and protocol analysis. These demands highlight the importance of integrating collaborative learning with mechanisms that reinforce individual accountability and reflection. A balanced approach that combines cooperative interaction with personal study routines and targeted feedback may better support the development of technical mastery.

While this study provides important insights into the academic and social dimensions of the Jigsaw method, it does not fully account for the psychological processes underpinning learning gains. Future work could incorporate complementary instruments to assess memory retention, metacognitive awareness, motivation, and self-regulated learning strategies. Such a multi-method approach would offer a more complete understanding of how cooperative learning environments influence educational outcomes.

The evidence presented supports the pedagogical value of the Aronson Jigsaw method. When implemented thoughtfully, it can promote sustained engagement, deepen learning, and improve student performance, especially when coupled with practices that enhance individual agency and conceptual accountability.

\section{Threats of Validity} \label{limitations}
This study provides meaningful insights into the pedagogical value of the Aronson method in technical subjects. However, several validity threats should be considered, along with the strategies used to reduce their impact.
\newline
\newline
\noindent
\textbf{Construct validity.} The Collaborative Learning Index (CLI) was designed as a composite indicator of engagement, combining scores from quizzes, group products, and peer or self evaluations. Although this approach includes multiple perspectives, it may not fully reflect deeper cognitive, metacognitive, or motivational processes. To reduce this limitation, the CLI was complemented with performance on final exams. Future studies should use validated instruments such as the Metacognitive Awareness Inventory or the Intrinsic Motivation Inventory to better capture these dimensions.
\newline
\newline
\noindent
\textbf{Internal validity.} Factors such as prior knowledge, learning strategies, or variations in instructional delivery may have influenced the results. While we did not directly control for these variables, the intervention was applied with the same structure in both teaching cases and delivered by the same instructor. This consistency helped reduce the variability caused by instructional style and content presentation.
\newline
\newline
\noindent
\textbf{External validity.} The small number of students and the specific university context may limit the general application of the findings. To address this concern, the study included two separate teaching cases from different subjects, showing that the technique is adaptable to distinct technical content. Future work should involve larger and more diverse student populations to strengthen generalization.
\newline
\newline
\noindent
\textbf{Conclusion validity.} Although appropriate statistical methods were used, the limited sample size could reduce the reliability of the conclusions. We used nonparametric tests to account for data distribution and sample size, and we ensured internal consistency of the CLI components. To further improve this aspect, future studies could use longitudinal data collection or resampling methods to reinforce the stability of the observed trends.

While these limitations exist, they were addressed through consistent study design, use of multiple data sources, and alignment of instructional conditions. Future research should build on this foundation by incorporating validated psychological scales, expanding sample diversity, and exploring artificial intelligence tools to support the analysis of student learning behaviors, especially in the teaching of design patterns and network protocols.

\section{Conclusion} \label{conclusion}
The implementation of the Aronson Jigsaw technique in two engineering education contexts—design patterns and computer networks—offered meaningful insights into the pedagogical impact of structured collaborative learning. Although no strong correlation was found between the Collaborative Learning Index and final individual performance, the cohorts exposed to the method demonstrated significantly improved academic results compared to previous groups. These findings suggest that the benefits of collaborative learning may be more accurately reflected in long-term conceptual gains rather than short-term indicators of engagement.

Pedagogically, the Aronson method appears to activate several key learning mechanisms. Through peer teaching and repeated content engagement, students benefit from distributed retrieval, which enhances memory consolidation. Preparing to teach promotes metacognitive processing, including planning and self-monitoring, while the collaborative setting increases motivation and encourages the development of self-regulated learning strategies. These cognitive, emotional, and behavioral processes likely contribute to better exam performance and deeper learning.

To capture these dimensions more fully, future studies should integrate additional assessment tools. For instance, concept retention could be measured using pre and post testing, while metacognitive engagement may be evaluated with validated self-report instruments or structured reflection. Motivation and self-regulation could be assessed through appropriate scales, and group dynamics analyzed using peer feedback and sociometric methods.

Looking ahead, future work could explore the integration of artificial intelligence tools specifically designed to support the teaching of design patterns and the OSI model within collaborative learning environments. For example, AI-based tutors could offer adaptive examples of pattern use or protocol behavior based on individual student weaknesses. Dialogue analysis tools powered by natural language processing could assess the conceptual depth of student discussions within expert groups, offering formative feedback to instructors. Additionally, AI-driven assessment systems could help evaluate the quality of peer teaching in real time, identifying conceptual gaps or misconceptions that emerge during collaborative sessions. These applications may enrich the Aronson framework by offering timely, data-informed interventions and deeper insights into how students construct and communicate technical knowledge.

While the present study offers a solid empirical foundation for evaluating the effectiveness of the Aronson method, a multi-method and AI-enhanced research agenda will be essential for uncovering the full scope of its cognitive and instructional potential in software engineering and network education.

\section*{conflict of interest}
The authors declare no conflict of interest.

\section*{Supporting Information}

dditional materials supporting the findings of this study, including anonymized data summaries, statistical analysis scripts, and implementation guidelines, are available at the following URL:

\printendnotes

\bibliography{bibliography}


\end{document}